\documentclass{scrartcl}

\usepackage{amsfonts}

\usepackage{subfigure}

\usepackage{tikz}
\usetikzlibrary{decorations.pathmorphing,decorations.markings}

\usepackage{url}

\title{Diffractive Dissociation into Three-Pion Final States at COMPASS}

\author{Sebastian Uhl\thanks{Speaker}~\thanks{The author acknowledges financial support by the organizers of the XV International Conference on Hadron Spectroscopy (Hadron 2013), by the German Bundesministerium f\"ur Bildung und Forschung (BMBF), by the Maier-Leibnitz-Laboratorium der Universit\"at und der Technischen Universit\"at M\"unchen, and by the DFG Cluster of Excellence ``Origin and Structure of the Universe'' (Exc153).}~~on behalf of the COMPASS Collaboration\\
Technische Universit\"at M\"unchen\\
E-mail: \url{sebastian.uhl@mytum.de}}

\date{}

\publishers{XV International Conference on Hadron Spectroscopy \\
		 4-8/11/2013\\
		 Nara, Japan}

\begin{document}

\maketitle

\begin{abstract}
In order to study diffractive dissociation reactions, COMPASS has taken data with a $190\,\textnormal{GeV}/c$ pion beam impinging on a liquid hydrogen target in 2008 and 2009. At squared four-momentum transfers to the target $t'$ between $0.1\,\textnormal{GeV}^2/c^2$ and $1.0\,\textnormal{GeV}^2/c^2$ the number of events with three pions in the final state is about an order of magnitude larger than that acquired by any previous experiment. In COMPASS, the three-pion final state can be studied in the two channels $\pi^-\pi^-\pi^+$ and $\pi^-\pi^0\pi^0$. The large data sample in particular for the $\pi^-\pi^-\pi^+$ channel allows to find even small signals at the sub-percent level. The progress of the partial-wave analysis will be shown. Compared to previous COMPASS results, the analysis is now performed in bins of $t'$, and the set of partial waves has been extended now including waves up to spin $6$. The information from the $t'$ dependence of the individual partial waves is very helpful in separating resonant and non-resonant contributions. As a consistency check results from the $\pi^-\pi^-\pi^+$ channel will be compared to the $\pi^-\pi^0\pi^0$ channel.
\end{abstract}

\section{Introduction}

One of the goals of the \textbf{Co}mmon \textbf{M}uon and \textbf{P}roton \textbf{A}pparatus for \textbf{S}tructure and \textbf{S}pectroscopy (COMPASS) experiment is to study the spectrum of light mesons. COMPASS is a fixed-target experiment located at CERN's Super Proton Synchrotron (SPS). The mesons under study are diffractively produced by a $190\,\textnormal{GeV}/c$ negative hadron beam, consisting mostly of pions with a small admixture of kaons and anti-protons, impinging on a liquid hydrogen target. The decay products of the diffractively produced mesons are detected in a two-stage magnetic spectrometer. Each of the two spectrometer stages is also equipped with an electromagnetic calorimeter. This setup provides full coverage for charged and neutral particles, resulting in a homogenous acceptance over a wide kinematic range. Events were recorded if a recoil proton was detected.

The reaction under study here is the diffractive production of mesons decaying into three pions, either into the $\pi^-\pi^-\pi^+$ or the $\pi^-\pi^0\pi^0$ final-state. To this end, exclusive events with a squared four-momentum transfer to the target $t'$ between $0.1\,\textnormal{GeV}/c^2$ and $1.0\,\textnormal{GeV}/c^2$ have been selected in a three-pion mass range between $0.5\,\textnormal{GeV}/c^2$ and $2.5\,\textnormal{GeV}/c^2$. This results in a huge dataset of $50$ million $\pi^-\pi^-\pi^+$ and $3.5$ million $\pi^-\pi^0\pi^0$ events.

Results from these two channels have already been presented before \cite{haas:2011, nerling:2011}. Compared to these references, the partial-wave analysis model applied to the data in the following has not only been extended in terms of the number of partial waves, but in addition the data are binned in the squared four-momentum transfer $t'$ in order to study the dependence of the production on this variable. Also a fit to the spin-density matrix of the $\pi^-\pi^-\pi^+$ channel has been performed \cite{paul:2013}.

\section{Partial-wave analysis}

\begin{figure}
\centering
\begin{tikzpicture}[thick,scale=0.75,
particle/.style={postaction={decorate}, decoration={markings,mark=at position .5 with {\arrow{>}}}},
pomeron/.style={postaction={decorate}, decoration={markings,mark=at position .5 with {\arrow{>}}}},
]
\draw[particle] (-2,1) -- (0,0);
\node[left] at (-2,1) {$\pi^-$};

\draw[particle] (0,0) -- node[above,sloped] {$X^-$} (2,0);

\draw[particle] (2,0) -- (6,1);
\draw[particle] (2,0) -- (4,-2);
\draw[particle] (4,-2) -- (6,-1);
\draw[particle] (4,-2) -- (6,-3);
\node[right] at (6,1) {$\pi$};
\node[right] at (6,-1) {$\pi$};
\node[right] at (6,-3) {$\pi$};

\draw[pomeron] (0,-2) -- node[right] {$\mathbb{P}$} (0,0);

\draw[particle] (-2,-3) -- (0,-2);
\draw[particle] (0,-2) -- (2,-3);
\node[left] at (-2,-3) {$p$};
\node[right] at (2,-3) {$p$};

\node[above,rectangle,fill=blue!20] at (1,2) {$J^{PC}M^\varepsilon$};
\path[->] (1,2) edge [bend right] (0.8,0.8);
\node[right,rectangle,fill=blue!20] at (3.5,-0.405) {$L$};
\path[ <-> ] (3,-1) edge [bend right] (3.372,0.343);
\node[below left,rectangle,fill=blue!20] at (2.875,-1.125) {$R_{\pi\pi}$};
\end{tikzpicture}
\caption{Sketch of the process under study.}
\label{fig:pwa.process}
\end{figure}
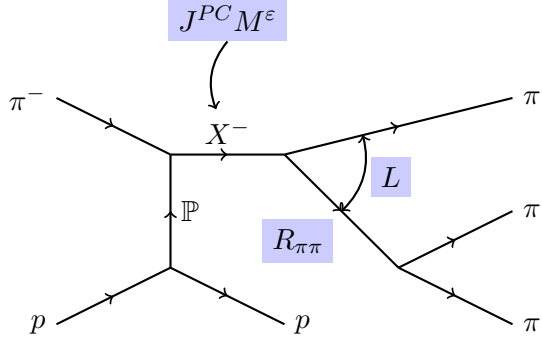

To decompose the three-pion mass spectra into the individual spin-parity components, a partial-wave analysis employing the isobar model is used. Fig.~\ref{fig:pwa.process} sketches the process under study. The incoming beam $\pi^-$ diffractively scatters off a target proton producing an intermediate state $X^-$, which subsequently decays into a two-pion isobar $R_{\pi\pi}$ and a bachelor pion which have a relative orbital angular momentum $L$. The two-pion isobar then decays into two pions. By exploiting the kinematic distribution in the five phase-space variables of the decay products, information on the spin $J$, the parity $P$, and the spin-projection $M^\varepsilon$ of $X^-$ can be extracted.

As isobars the $\rho\left(770\right)$, $f_0\left(980\right)$, $f_2\left(1270\right)$, $f_0\left(1500\right)$, $\rho_3\left(1690\right)$, and a broad $\left(\pi\pi\right)_S$ component are used. Partial waves with a spin $J$ up to $6$ are included, also the angular momentum $L$ between the isobar $R_{\pi\pi}$ and the bachelor pion can go up to $6$. Out of the possible combinations, $87$ waves with non-negligible intensity are kept for the final analysis: $80$ waves with a positive reflectivity $\varepsilon=+1$ and $7$ with $\varepsilon=-1$. In addition one incoherent wave with an isotropic angular distribution is included.

\section{Fit in mass bins}

The decomposition of the data into partial waves is performed with two different programs, which have been cross-checked versus each other. The fit for the $\pi^-\pi^-\pi^+$ is performed using a program originally developed at Illinois \cite{ascoli:1970}, which was later modified in Protvino and Munich. For the fit to the $\pi^-\pi^0\pi^0$ channel ROOTPWA is used \cite{rootpwa:2013} (originally based on \cite{cummings:2003}).

\begin{figure}
\centering
\subfigure[$2^{++}1^+\rho\left(770\right) \pi D$]{\includegraphics[height=5.5cm]{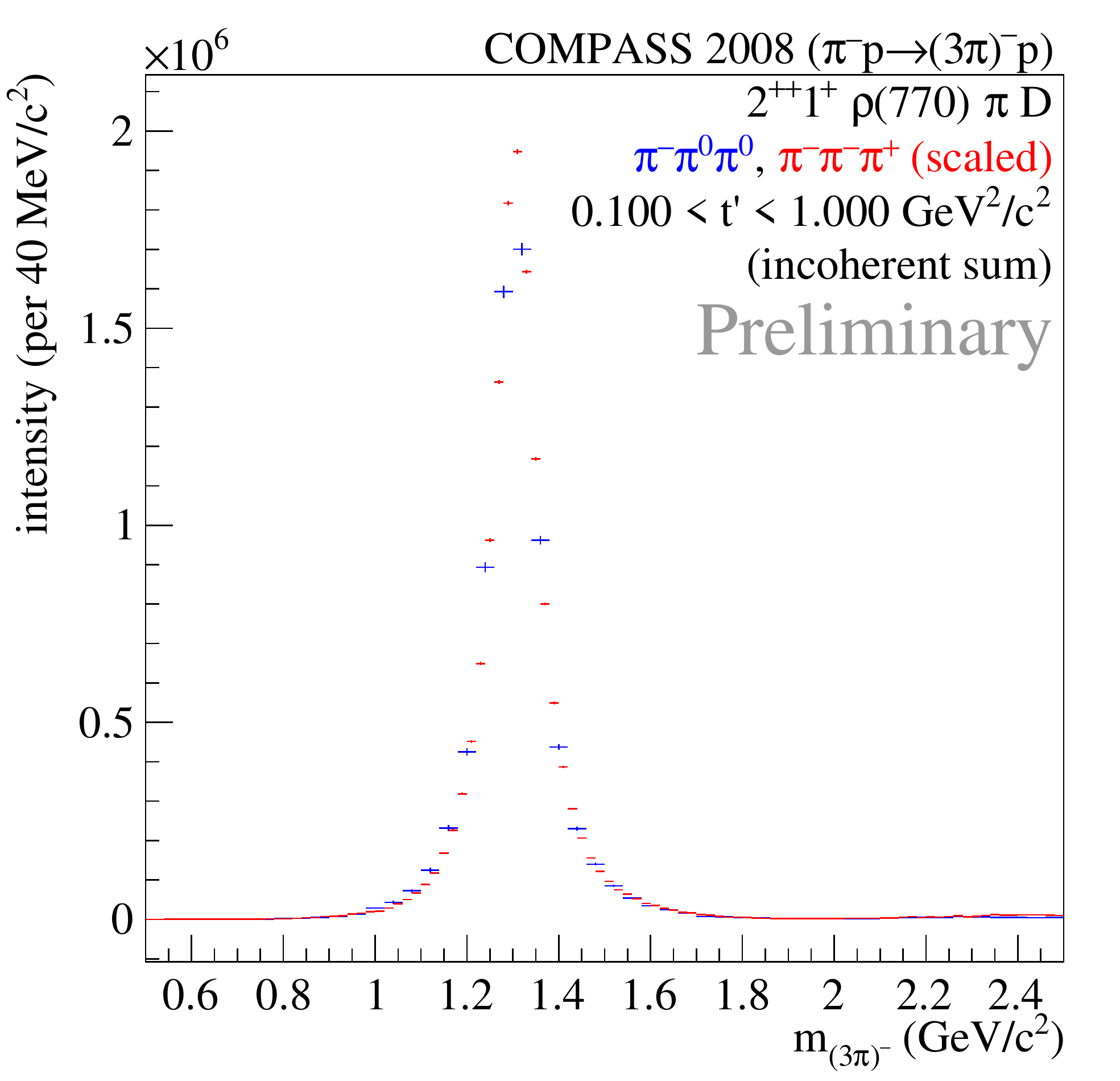}}
\subfigure[$2^{-+}0^+f_2\left(1270\right) \pi S$]{\includegraphics[height=5.5cm]{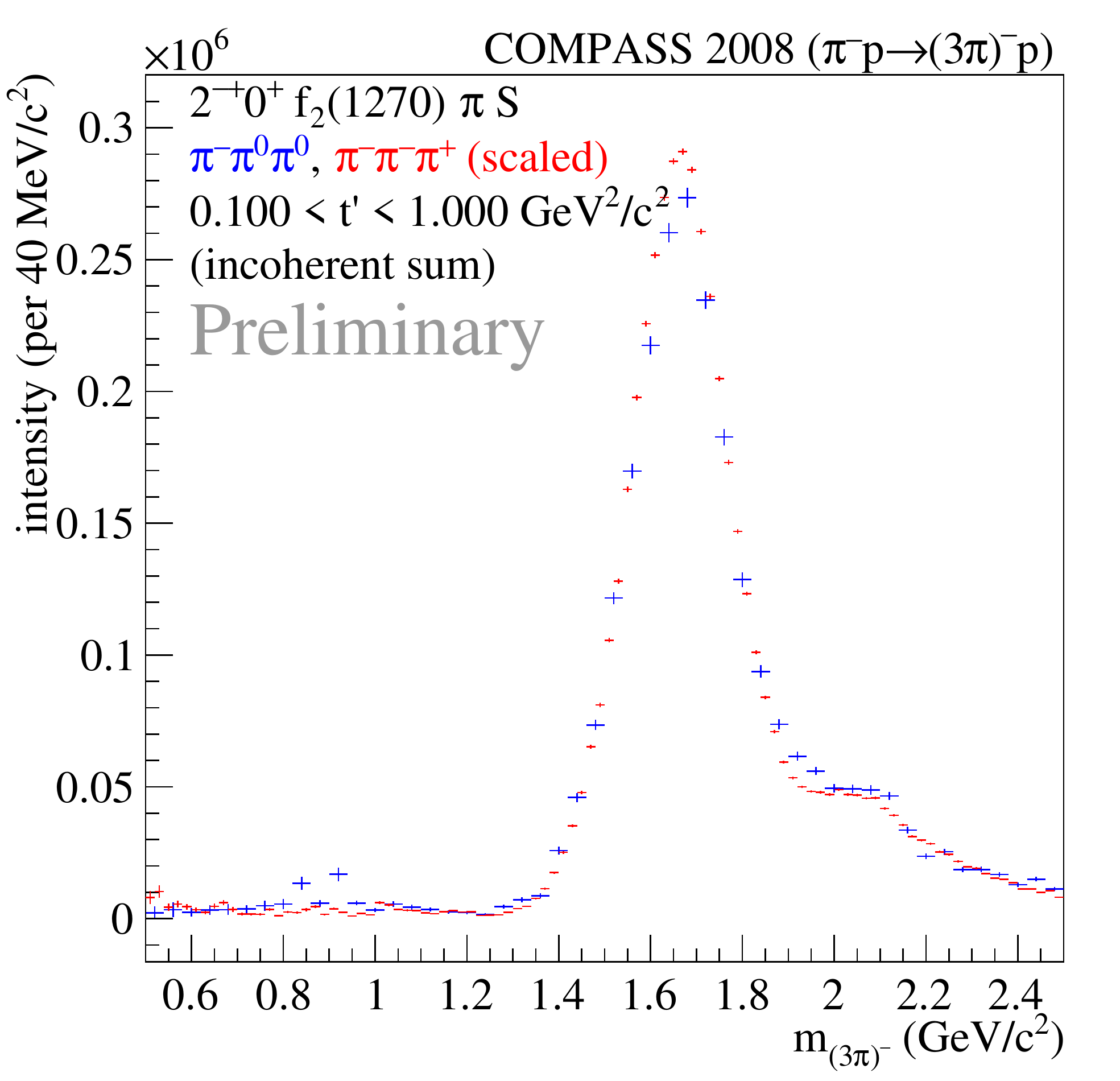}}
\caption{Incoherent sum of partial-wave intensities over all $t'$ bins.}
\label{fig:pwa.a2pi2}
\end{figure}

\begin{figure}
\centering
\subfigure[low $t'$]{\includegraphics[height=5.5cm]{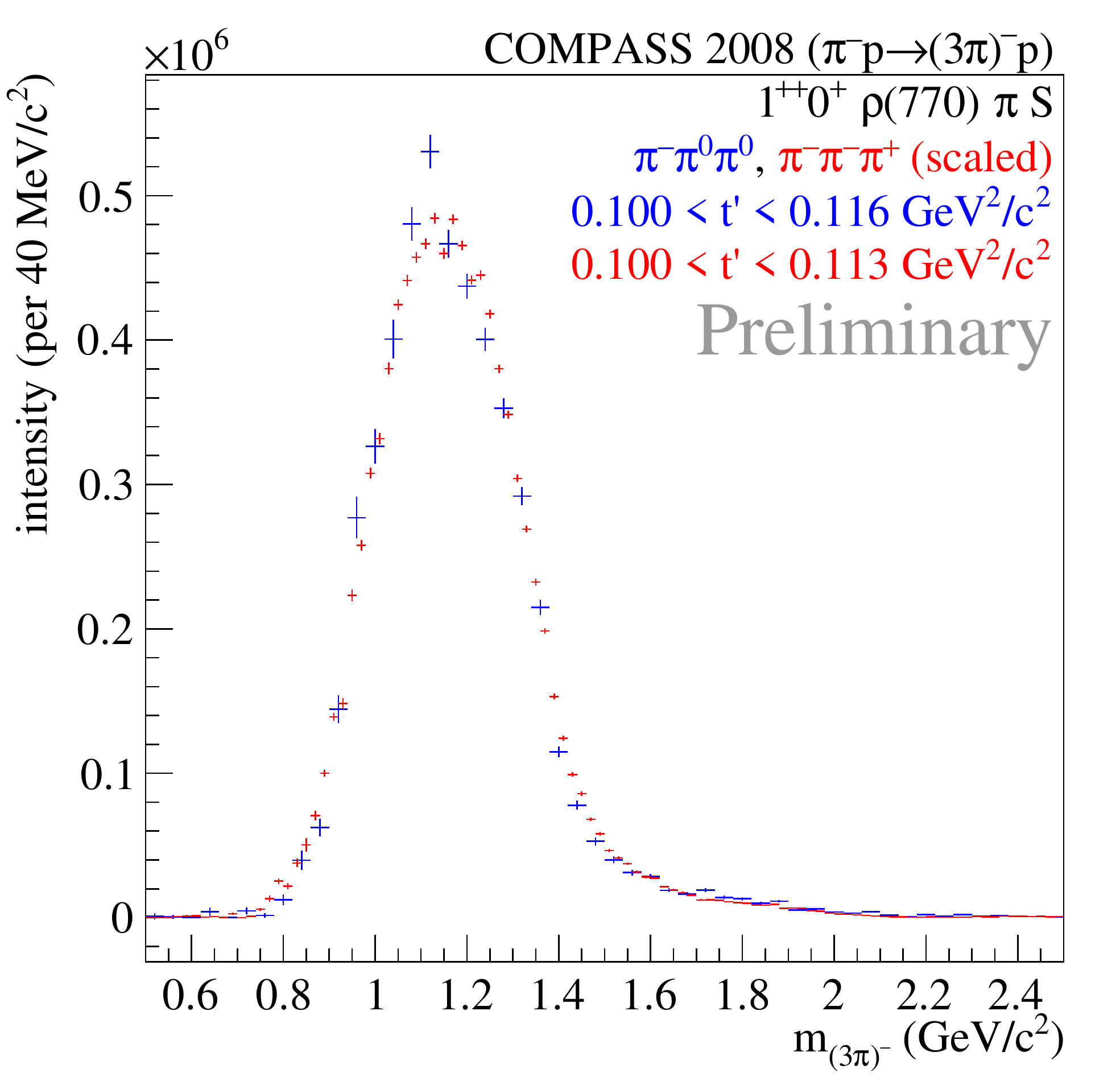}}
\subfigure[high $t'$]{\includegraphics[height=5.5cm]{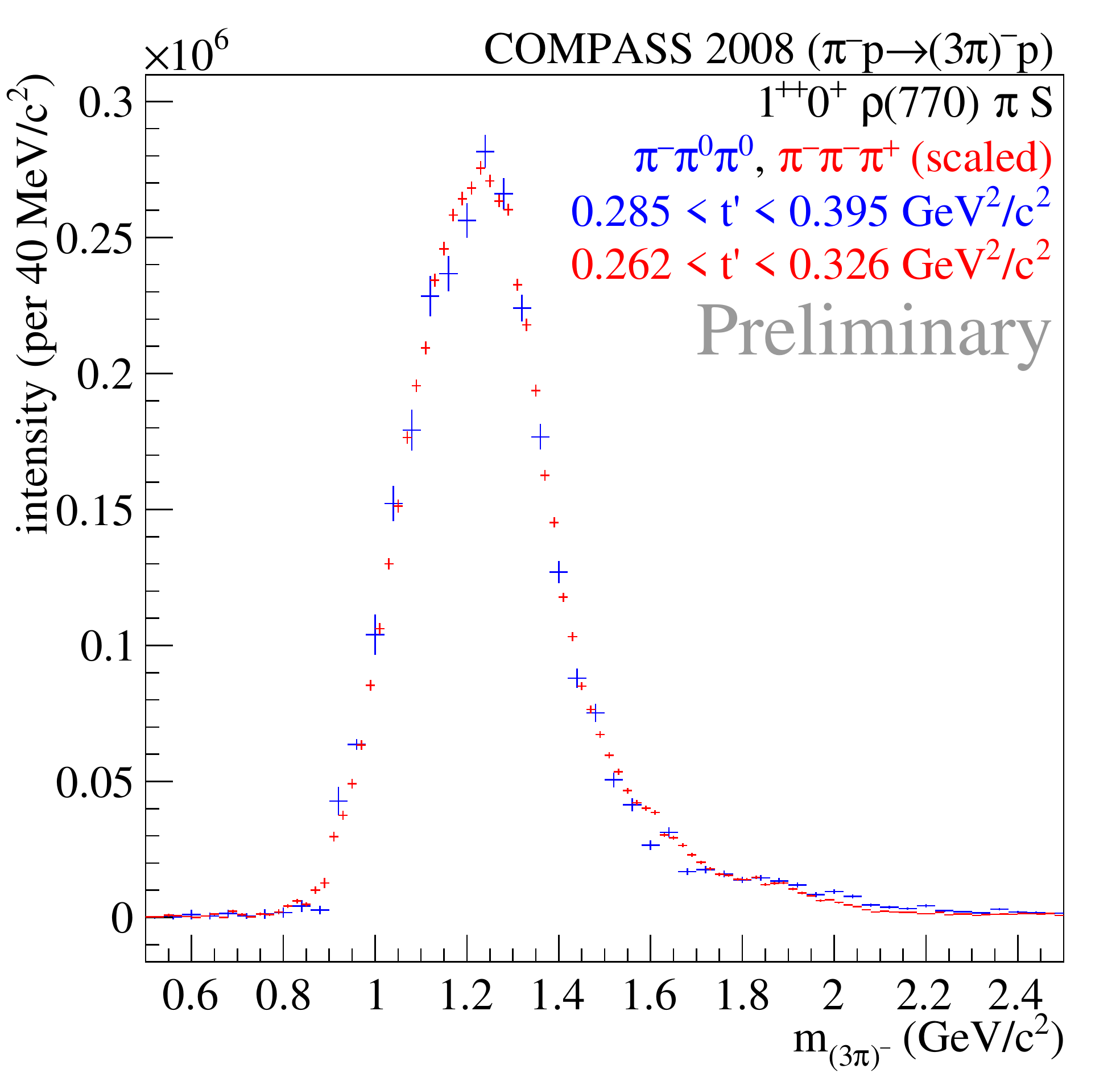}}
\caption{Intensities of the $1^{++}0^+\rho\left(770\right) \pi S$ wave in two $t'$ regions.}
\label{fig:pwa.a1}
\end{figure}

A rank-1 fit is performed in bins of both the three-pion mass and the squared four-momentum transfer to the target $t'$. Fig.~\ref{fig:pwa.a2pi2} shows the incoherent sum of the intensities over the individual $t'$ bins for the $2^{++}1^+\rho\left(770\right) \pi D$ wave, clearly showing the $a_2\left(1320\right)$, and the $2^{-+}0^+f_2\left(1270\right) \pi S$ wave, clearly showing the $\pi_2\left(1670\right)$. The results from the $\pi^-\pi^-\pi^+$ channel (blue markers) and the $\pi^-\pi^0\pi^0$ channel (red markers) are in agreement. The two data sets are normalized using the intensity integrals in each individual plot so that the peak shapes can directly be compared. In contrast to those two waves, which show the same shape also in the individual $t'$ bins, the shape of the $1^{++}0^+\rho\left(770\right) \pi S$ wave, where the $a_1\left(1260\right)$ is expected, changes with $t'$ (Fig.~\ref{fig:pwa.a1}) indicating a rather large non-resonant contribution.

\begin{figure}
\centering
\subfigure[incoherent sum over all $t'$ bins\label{fig:pwa.a11420.a}]{\includegraphics[height=5.5cm]{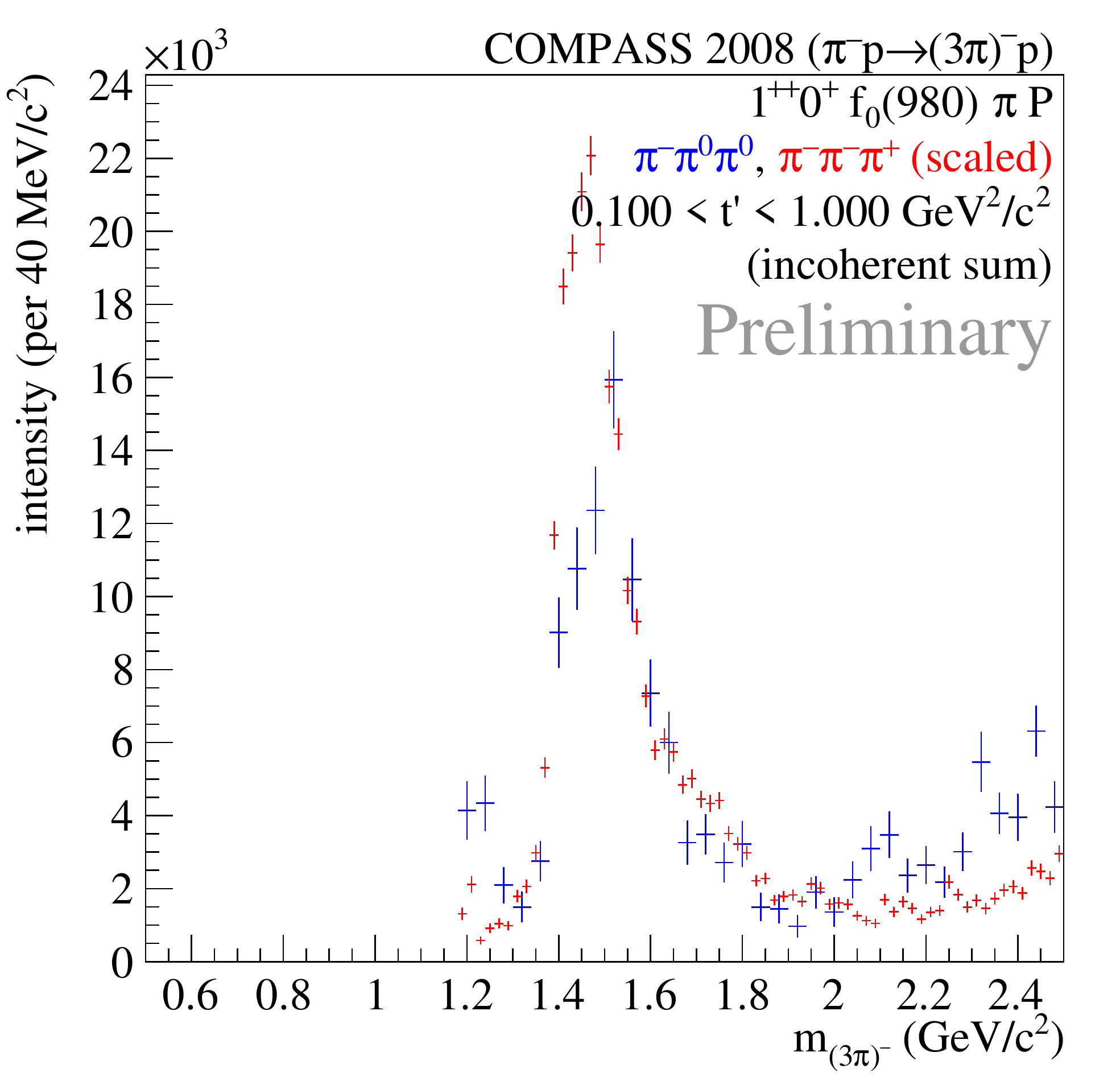}}
\subfigure[fit to spin-density matrix\label{fig:pwa.a11420.b}]{\begin{tikzpicture}[path image/.style={
      path picture={
        \node at (path picture bounding box.center) {
          \includegraphics[height=5.5cm, trim=13.25cm 0cm 0cm 15.2cm, clip=true]{#1}
        };}}
]
\draw [path image=intensity_wave_5, draw=white] (0,0) rectangle (8.cm,5.5cm);

\draw (3.2cm,2.2cm) node[rotate=34.5085,scale=2.,text=gray,text opacity=10] {Preliminary};
\end{tikzpicture}}
\caption{Intensities of the $1^{++}0^+f_0\left(980\right) \pi P$ wave.}
\label{fig:pwa.a11420}
\end{figure}
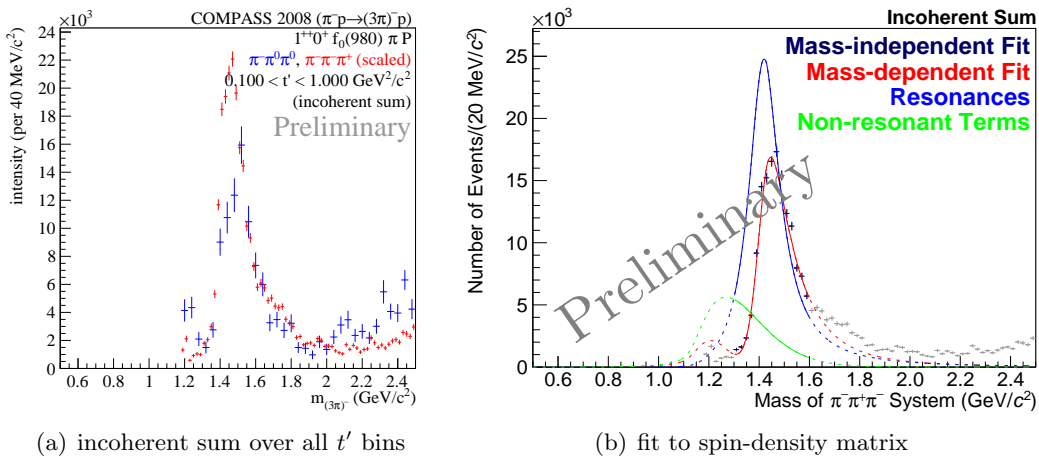

In addition to the intensities and phase motions of a number of smaller waves that are in good agreement with resonances previously observed, one particularly interesting signal is observed in the $1^{++}0^+f_0\left(980\right) \pi P$ wave (Fig.~\ref{fig:pwa.a11420.a}) around $1.4\,\textnormal{GeV}/c^2$. This narrow structure, never observed before, could correspond to a possible new $a_1$ state. It is seen in both channels and for all $t'$ bins. Different parameterizations for the isobars have been tested to exclude a possible artifact from the used model.

\section{Fit to spin-density matrix}

To further study this new signal, and to extract Breit-Wigner parameters, the mass dependence of a sub-set of the spin-density matrix was extracted for the $\pi^-\pi^-\pi^+$ channel. Apart from the intensities, also the interferences between the waves were taken into account. In addition to the four waves mentioned above, also the $4^{++}1^+\rho\left(770\right) \pi G$ and the $0^{-+}0^+f_0\left(980\right) \pi S$ partial waves were included. The fit model included the following Breit-Wigner resonances: the $a_1\left(1260\right)$, an $a_1'$, the $a_2\left(1320\right)$, an $a_2'$, the $a_4\left(2040\right)$, $\pi\left(1800\right)$, $\pi_2\left(1670\right)$, $\pi_2\left(1880\right)$, and a new $a_1\left(1420\right)$ in the $1^{++}0^+f_0\left(980\right) \pi P$ wave. The Breit-Wigner parameters were extracted from a simultaneous fit to the spin-density submatrices of the six partial waves in all eleven $t'$ bins. Apart from the Breit-Wigner components used to describe the resonances, also a coherent $t'$-dependent non-resonant contribution was allowed in each wave. The parameters of the Breit-Wigner functions and the coherent background were the same in all $t'$ bins, only the complex couplings were allowed to differ.

The extracted parameters of the major resonances are in agreement with previous measurements by COMPASS and other experiments \cite{compass:2010}. Fig.~\ref{fig:pwa.a11420.b} shows the result from this fit for the new $a_1\left(1420\right)$. The intensity extracted by the spin-parity decomposition in three-pion mass bins (black markers) is well described by the model (red line), consisting of the resonance part (blue line) and a non-resonant contribution (green line). A well determined mass $M=1412-1422\,\textnormal{MeV}/c^2$ and width $\Gamma=130-150\,\textnormal{MeV}/c^2$ was obtained for the new $a_1\left(1420\right)$. This is in contrast to other small signals like the $a_1'$, the $a_2'$, or the $\pi_2\left(1880\right)$ for which a large uncertainty of the mass and width is retrieved from the same fit.

In order to exclude possible artifacts caused by the particular choice for the parametrization of the isoscalar $J^{PC}=0^{++}$ isobars, a partial-wave analysis was performed, in which the waves with $\left(\pi\pi\right)_S$, $f_0\left(980\right)$, and $f_0\left(1500\right)$ isobars are replaced by amplitudes piecewise constant in the two-pion mass \cite{paul:2013}. This new approach allows a rather model-independent extraction of the $0^{++}$ isobar amplitudes. The analysis shows a clear correlation between the new $a_1\left(1420\right)$ signal and the $f_0\left(980\right)$ as a decay product.

\section{Conclusions}

The COMPASS experiment has recorded a huge dataset to study resonances produced in diffractive dissociation into three pions. The large number of events allowed to increase the number of waves used in the partial-wave analysis, while still being able to also perform a $t'$-dependent analysis. Breit-Wigner parameters have been extracted from a fit to the result of the spin-parity decomposition in mass bins. Apart from the well-known resonances, COMPASS found a new signal in the $1^{++}0^+f_0\left(980\right) \pi P$ wave around $1.4\,\textnormal{GeV}/c^2$. It is a rather narrow object featuring a well-defined mass and width. Its nature is still unclear, it could be the isospin partner of the $f_1\left(1420\right)$, but also other explanations like rescattering effects \cite{basdevant:1977} cannot be ruled out at this stage of the analysis.


\begin{thebibliography}{99}
  \bibitem{haas:2011} F.~Haas on behalf of the COMPASS Collaboration, in Proceedings of the XIV International Conference on Hadron Spectroscopy (hadron2011), Munich, 2011, edited by B.~Grube, S.~Paul, and N.~Brambilla, eConf C110613 (2011) 1109.1789.
  \bibitem{nerling:2011} F.~Nerling on behalf of the COMPASS Collaboration, in Proceedings of the XIV International Conference on Hadron Spectroscopy (hadron2011), Munich, 2011, edited by B.~Grube, S.~Paul, and N.~Brambilla, eConf C110613 (2011) 1108.5969.
  \bibitem{paul:2013} S.~Paul on behalf of the COMPASS Collaboration, proceedings of the MENU 2013, arXiv:1312.3678.
  \bibitem{ascoli:1970} G.~Ascoli \textit{et al.}, Phys. Rev. Lett. \textbf{25}, 962 (1970).
  \bibitem{rootpwa:2013} ROOTPWA, URL \url{http://sourceforge.net/projects/rootpwa}.
  \bibitem{cummings:2003} J.P.~Cummings, D.P.~Weygand, arXiv:physics/0309052.
  \bibitem{compass:2010} COMPASS collaboration, M.G.~Alekseev \textit{et al.}, Phys. Rev. Lett. \textbf{104}, 241803 (2010).
  \bibitem{basdevant:1977} J.L.~Basdevant, E.~Berger, Phys. Rev. D\textbf{16}, 657 (1977).
\end{thebibliography}
\end{document}